# Effect of impurities on the transition temperature of a dilute dipolar tapped Bose gase

H. Yavari*, E. Afsaneh
Department of Physics, University of Isfahan, Hezar Jarib, Isfahan 81746, Iran

**Abstract:**
By using two-fluid model the effect of impurities on the transition temperature of a dipolar trapped Bose gas is investigated. By treating Gaussian spatial correlation for impurities from the interaction modified spectra of the system, the formula for the shift of transition temperature is derived. The shift of the transition temperature contains essentially three contributions due to contact, dipole-dipole and impurity interactions. Applying our results to Dipolar Bose gases show that the shift of the transition temperature due to impurities could be measured for isotropic trap (dipole-dipole contributions is zero) and Feshbach resonance technique (contact potential contribution is negligible).

## 1-Intruduction

The study of ultracold atomic systems has received much attention in recent years, motivated largely by the prospect of realizing novel strongly correlated many-body physics, in which macroscopic quantum phenomena can be studied experimentally over a wide range of controllable interactions [1–3]. For the original alkali atomic Bose-Einstein condensates (BECs) it has been sufficient to describe the dominant two-particle interaction by a local isotropic contact potential. Recently, a new type of nonlocal anisotropic interaction has been made accessible to detailed study by the formation of a BEC in a dipolar quantum gas of $^{52}$Cr [4], where the magnetic dipole moments are around six times larger than those of alkali-metal atoms. Therefore, the physical properties of such a chromium BEC also depend on a long range and anisotropic magnetic dipole-dipole interaction. The atom-atom interaction is then determined by the balance of both potentials, giving rise the interesting phenomena in a dipole Bose Einstein condensation. Other many-body systems with dipolar interactions are, for instance, Rydberg atoms [5, 6] or atomic condensates where a strong electric field induces electric dipole moments of the order of $10^{-2}$ D [7].

The use of Feshbach resonances to tune the contact interaction has opened new possibilities of control over such systems. By applying an external magnetic field to tune the s-wave scattering length it is possible to explore extreme regimes from strongly repulsive to very small and strongly attractive interaction. One prominent example called 'Bosenova' is the collapse and explosion of a BEC when the contact interaction is suddenly changed from repulsive to attractive [8]. Feshbach resonances have also been used to create BECs of bosonic molecules formed by fermionic atoms [9–11] and to study the BEC-BCS crossover in these systems [12–16].

It is evident that creating a Bose–Einstein condensate of atoms that are subject to an interaction with different symmetry and range than the contact interaction will also introduce new collective phenomena. It has been shown in [7, 17, and 18] that the anisotropy of the magnetic dipole–dipole interaction (MDDI) introduces anisotropy in the density distribution of a trapped dipolar condensate. The experimental observation of a modification of the condensate expansion due to the MDDI that was proposed in [19] is a central point of this tutorial.

The regime of dominant dipole–dipole interaction can be entered by using a Feshbach resonance to lower the contact interaction. Moreover, a continuous transition between both regimes of dominant dipole–dipole or contact interaction is possible.

A wide variety of different (but related) approaches to the theory of BEC were developed for the homogeneous case and have now been extended to trapped gases. The first quantitative analysis was given by Bogoliubov in 1947 [20] and was based on approximating the Hamiltonian of the system by one which is quadratic and hence can be diagonalized exactly. The extension of this method to higher-order calculations and trapped gases has become known as the Hartree– Fock–Bogoliubov (HFB) theory [21]. Bogoliubov obtained his quadratic Hamiltonian by using a description of BEC in terms of spontaneous symmetry breaking.

Much of the theoretical research on trapped Bose gases in recent years has been on dynamical issues, e.g., condensate formation, damping of collective excitations, and collapse of the condensate. There is also a large body of literature on the nonequilibrium dynamics of the dilute Bose gas.

The critical temperature is the temperature at which all the particles can be accommodated in the excited states. For an ideal Bose gas, this is in the two-thirds power of the number density $T_c^0 = 4\pi \left[ \dfrac{N}{\zeta\left(\frac{3}{2}\right)} \right]^{2/3}$, and the number of condensate atoms below $T_c^0$ is in the 1.5th power of the temperature. With the trap potential, the corresponding $T_c^0$ changed into the one-third power of the number of atoms $T_c^0 = \dfrac{\hbar\omega}{k_B} \left[ \dfrac{N}{\zeta(3)} \right]^{1/3}$, and number of condensate atoms for temperature below $T_c^0$ is to the third power of the temperature [22].

For the homogenous Bose gas one might think that the effect of a repulsive interaction is to decrease the critical temperature of a Bose gas. For instance, the superfluid transition in liquid $^4$He takes place at a lower temperature than that of an ideal gas of the same density. However, liquid $^4$He is not weakly interacting, and it turns out that the leading effect of the interactions in the dilute Bose gas is to increase $T_c$. In contrast to that, for a trapped Bose gas trapping potentials remove the critical long-wavelength fluctuations and reduce the fraction of atoms taking part in nonperturbative physics at the transition point. As a result, the leading shift of the transition temperature with respect to its noninteracting value is negative.

The study of the effects of particle interactions on the critical temperature of the Bose-Einstein condensation has a long and tortuous history. Several studies on the homogeneous Bose gases found that repulsive interactions decrease the critical temperature [23-25], but most studies pointed to an increase of the critical temperature [26-34].

Besides the contribution of the well-known atomic contact potential term on the transition temperature, there is an additional contribution from dipole-dipole interaction. The effects of these two interactions on the transition temperature ($T_c$) of a trapped Bose gas was investigated by using the two-fluid model [35, 36].

Apart from the contact and dipole-dipole interactions, another interesting physical parameter for a trapped Bose gas is impurities. It can be created artificially by laser speckles [37], incommensurate lattices [38], or different localized atomic species [39, 40]. Thus, one can raise the fundamental question how the Bose-Einstein condensation temperature is affected by impurity.
In this paper by following the method of ref [35], we investigate how the critical temperature of a dipolar bose gas depends on the dipole-dipole, contact, and impurity interactions.

**2-Formalism**
The starting point is the second quantized grand-canonical Hamiltonian for interacting bosons of mass m that includes two-body and impurity potentials

$$H = \int d^3r \left[ \hat{\psi}^\dagger(\vec{r},t) \left( -\frac{\hbar^2 \nabla^2}{2m} + U_{ext}(\vec{r}) - \mu \right) \hat{\psi}(\vec{r},t) \right.$$

$$\left. + \int d^3r d^3r' \hat{\psi}^\dagger(\vec{r}',t) \hat{\psi}^\dagger(\vec{r},t) U(\vec{r},\vec{r}') \hat{\psi}(\vec{r}',t) \hat{\psi}(\vec{r},t) \right] \quad (1)$$

$$+ \int d^3r d^3r' \psi^\dagger(\vec{r}') \varphi^\dagger(\vec{r}) U_{imp}(\vec{r},\vec{r}') \varphi(\vec{r}') \psi(\vec{r})$$

where $\hat{\psi}(\vec{r},t)$ and $\hat{\psi}^\dagger(\vec{r},t)$ are bosonic annihilation and creation fields, , $\varphi^\dagger$ and $\varphi$ are field operators for the impurity, $U_{ext}(\vec{r})$ is the confining potential, $U(\vec{r},\vec{r}')$, the interaction potential including two-body interaction, and $U_{imp}(\vec{r},\vec{r}')$, impurity potential can be written respectivel as

$$U(\vec{r},\vec{r}') = U_{contact}(\vec{r},\vec{r}') + U_{dipole-dipole}(\vec{r},\vec{r}')$$
$$= g\delta(\vec{r}-\vec{r}') + g_d U_d(\vec{r},\vec{r}') \quad (2)$$

$$U_{imp}(\vec{r},\vec{r}') = g_{iB} \delta(\vec{r}-\vec{r}') \quad (3)$$

here the coupling constant g is related to the s-wave scattering length $a$ by $g = \frac{4\pi a \hbar^2}{m}$, $g_d = \frac{\mu_0 \mu_{Cr}^2}{4\pi}$ and $U_d(\vec{r},\vec{r}') = \frac{1-3\cos^2\theta}{|\vec{r}-\vec{r}'|^3}$ describe the dipole-dipole interaction potential between two dipoles. The atomic magnetic dipole moments $\mu_{Cr}$ are aligned, and $\theta$ is the angle between the relative position vector $\vec{r}-\vec{r}'$. The coupling constants $g_{iB}$ describes the local interaction between bosons and i-th impurity is related to the corresponding s-wave scattering lengths $a_{iB}$ and reduced masses $m_r = \frac{m_i m_B}{m_i + m_B}$ by $g_{iB} = \frac{2\pi a_{iB} \hbar^2}{m_r}$,
By using Eqs. (2) and (3) into Eq. (1) we obtain

$$H = \int d^3r \left[ \hat{\psi}^\dagger(\vec{r},t) \left( -\frac{\hbar^2 \nabla^2}{2m} + U_{ext}(\vec{r}) - \mu \right) \hat{\psi}(\vec{r},t) \right.$$

$$\left. + \frac{g}{2} \hat{\psi}^\dagger(\vec{r},t) \hat{\psi}^\dagger(\vec{r},t) \hat{\psi}(\vec{r},t) \hat{\psi}(\vec{r},t) \right] \quad (4)$$

$$+ g_d \int d^3r \, d^3\vec{r}' \hat{\psi}^\dagger(\vec{r}',t) \hat{\psi}^\dagger(\vec{r},t) U_d(\vec{r},\vec{r}') \hat{\psi}(\vec{r}',t) \hat{\psi}(\vec{r},t)$$

$$+ g_{iB} \int d^3r \hat{\psi}^\dagger(\vec{r},t) \varphi^\dagger(\vec{r},t) \varphi(\vec{r},t) \hat{\psi}(\vec{r},t)$$

The usual Heisenberg equation of motion for the quantum field operator is

$$i\hbar \frac{\partial \hat{\psi}(\vec{r},t)}{\partial t} = \left[ \hat{\psi}(\vec{r},t), H \right] = \left( -\frac{\hbar^2 \nabla^2}{2m} + U_{ext}(\vec{r}) \right) \hat{\psi}(\vec{r},t) + g \hat{\psi}^\dagger(\vec{r},t) \hat{\psi}(\vec{r},t) \hat{\psi}(\vec{r},t)$$

$$+ g_d \left( \int d^3\vec{r}' \hat{\psi}^\dagger(\vec{r}',t) \hat{\psi}(\vec{r}',t) U_d(\vec{r},\vec{r}') \right) \hat{\psi}(\vec{r},t) + g_{iB} \varphi^\dagger(\vec{r},t) \varphi(\vec{r},t) \hat{\psi}(\vec{r},t) \quad (5)$$

Separating out the condensate part in the usual way, we have

$$\hat{\psi}(\vec{r},t) = \Phi(\vec{r}) + \psi(\vec{r},t) \quad (6)$$

where the averaged $\Phi(\vec{r}) = \langle \hat{\psi}(\vec{r},t) \rangle$ plays the role of a spatially varying macroscopic Bose field and $\psi(\vec{r},t)$ is the non-condensate field operator.

with $\langle \psi(\vec{r},t) \rangle = 0$, the expectation value of (5) thus yields

$$i\hbar \frac{\partial \Phi(\vec{r},t)}{\partial t} = \left( -\frac{\hbar^2 \nabla^2}{2m} + U_{ext}(\vec{r}) - \mu \right) \Phi(\vec{r}) + g \left\langle \hat{\psi}^\dagger(\vec{r},t) \hat{\psi}(\vec{r},t) \hat{\psi}(\vec{r},t) \right\rangle$$

$$+ g_d \left( \int d^3\vec{r}' \left\langle \hat{\psi}^\dagger(\vec{r}',t) \hat{\psi}(\vec{r}',t) \hat{\psi}(\vec{r},t) \right\rangle U_d(\vec{r},\vec{r}') \right) + g_{iB} |\varphi|^2 \Phi(\vec{r}) \quad (7)$$

Follow the method of Griffin, the field operator part in the interaction terms of Eq. (7) can be written as

$$\hat{\psi}^\dagger(\vec{r}',t) \hat{\psi}(\vec{r}',t) \hat{\psi}(\vec{r},t) = |\Phi(\vec{r}')|^2 \Phi(\vec{r}) + |\Phi(\vec{r}')|^2 \psi(\vec{r},t)$$

$$+ \Phi(\vec{r}') \psi(\vec{r}',t) \Phi(\vec{r}) + \Phi(\vec{r}') \psi^\dagger(\vec{r}',t) \Phi(\vec{r}) + \Phi^*(\vec{r}') \psi(\vec{r}',t) \psi(\vec{r},t) \quad (8)$$

$$+ |\hat{\psi}(\vec{r}',t)|^2 \Phi(\vec{r}) + \psi^\dagger(\vec{r}',t) \Phi(\vec{r}') \psi(\vec{r},t) + |\psi(\vec{r}',t)|^2 \psi(\vec{r},t)$$

Introduce the local densities we obtain the Gross-Pitaevskii equation (GPE) for the condensate part:

$$i\hbar \frac{\partial \Phi(\vec{r},t)}{\partial t} = \left( -\frac{\hbar^2 \nabla^2}{2m} + U_{ext}(\vec{r}) - \mu \right) \Phi(\vec{r})$$

$$+ g_d \left( \int d^3r' \left[ n_c(\vec{r}') + n_T(\vec{r}') \right] U_d(\vec{r},\vec{r}') \right) \Phi(\vec{r})$$

$$g_d \left( \int d^3r \, n_T(\vec{r},\vec{r}') U_d(\vec{r}',\vec{r}) \Phi(\vec{r}') \right) + g_d \left( \int d^3r \, m_T(\vec{r}',\vec{r}) U_d(\vec{r},\vec{r}') \Phi^*(\vec{r}') \right) \quad (9)$$

$$+ g \left( n_c(\vec{r}) + 2n_T(\vec{r}) \right) \Phi(\vec{r}) + g m_T(\vec{r}) \Phi^*(\vec{r})$$

$$+ g \left\langle \psi^\dagger(\vec{r},t) \psi(\vec{r},t) \psi(\vec{r},t) \right\rangle + g_{iB} |\varphi|^2 \Phi(\vec{r})$$

where $n_c(\vec{r},t) = |\Phi(\vec{r},t)|^2$ is the non-equilibrium density of atoms in the condensate and $n_T(\vec{r},t) = \langle \psi^\dagger(\vec{r},t)\psi(\vec{r},t) \rangle$ is the non-equilibrium non-condensate density. Equation (9) also involves the off-diagonal non-condensate density $m(\vec{r},t) = \langle \psi(\vec{r},t)\psi(\vec{r},t) \rangle$ and the three-field correlation function $\langle \psi^\dagger \psi \psi \rangle$, both of which in principle have non-zero expectation values because of the assumed Bose broken symmetry. It is useful at this point to discuss some standard approximations to (9). In the absence of dipole-dipole interaction and impurities it reduces to the well-known Gross-Pitaevskii (GP) equation [23] if all the atoms are in the condensate (i.e., $n_T = 0$) and the anomalous correlations $m$ and $\langle \psi^\dagger \psi \psi \rangle$ are ignored. This is a very good approximation for $T \ll T_{BEC}$: at $T = 0$, the non-condensate fraction in trapped atomic gases is estimated to be less than 1%.[41, 42]. The excitations of the condensate are described by $\psi(\vec{r},t)$ and are given by subtraction of the time average part of Eq. (5):

$$i\hbar \frac{\partial \psi(\vec{r},t)}{\partial t} = \left( -\frac{\hbar^2 \nabla^2}{2m} + U_{ext}(\vec{r}) \right) \psi(\vec{r},t) + g n_c \psi^\dagger(\vec{r},t) - 2g n_T \psi(\vec{r},t)$$

$$+ g \Phi^*(\psi\psi - m) + 2g \Phi(\psi^\dagger \psi - n_T) + g \left( \psi^\dagger \psi \psi - \langle \psi^\dagger \psi \psi \rangle \right)$$

$$+ g_d \int d^3 \vec{r}' \left[ \hat{\psi}^\dagger(\vec{r}',t) \hat{\psi}(\vec{r}',t) \psi(\vec{r},t) U_d(\vec{r},\vec{r}') \right]$$

$$- g_d \int d^3 \vec{r}' \left[ \langle \hat{\psi}^\dagger(\vec{r}',t) \hat{\psi}(\vec{r}',t) \psi(\vec{r},t) \rangle U_d(\vec{r},\vec{r}') \right] + g_{iB} |\varphi|^2 \psi(\vec{r},t)$$

(10)

Under the mean-field approximation

$$\psi^\dagger(\vec{r}',t)\psi(\vec{r},t) \approx \langle \psi^\dagger(\vec{r}')\psi(\vec{r}) \rangle = n_T(\vec{r}',\vec{r}) \tag{11}$$

$$\psi(\vec{r}',t)\psi(\vec{r},t) \approx \langle \psi(\vec{r}')\psi(\vec{r}) \rangle = m_T(\vec{r}',\vec{r}) \tag{12}$$

and

$$\hat{\psi}^\dagger(\vec{r}',t)\hat{\psi}(\vec{r}',t)\hat{\psi}(\vec{r},t) - \langle \hat{\psi}^\dagger(\vec{r}')\hat{\psi}(\vec{r}')\hat{\psi}(\vec{r}) \rangle$$

$$\approx \langle \hat{\psi}^\dagger(\vec{r}')\hat{\psi}(\vec{r}') \rangle \psi(\vec{r},t) + \langle \hat{\psi}^\dagger(\vec{r}')\hat{\psi}(\vec{r}) \rangle \psi(\vec{r}',t) + \langle \hat{\psi}(\vec{r}')\hat{\psi}(\vec{r}) \rangle \psi(\vec{r}',t)$$

(13)

Define the self-consistent densities

$$n(\vec{r}) = \langle \hat{\psi}^\dagger(\vec{r},t)\hat{\psi}(\vec{r},t) \rangle = n_c(\vec{r}) + n_T(\vec{r}) \tag{14}$$

$$m(\vec{r}) = \langle \hat{\psi}(\vec{r},t)\hat{\psi}(\vec{r},t) \rangle = \Phi^2(\vec{r}) + m_T(\vec{r}) \tag{15}$$

Then, Eq. (10) reduces to

$$i\hbar \frac{\partial \psi(\vec{r},t)}{\partial t} = \left( -\frac{\hbar^2 \nabla^2}{2m} + U_{ext}(\vec{r}) \right) \psi(\vec{r},t) + g n_c \psi^\dagger(\vec{r},t) + 2g n_T \psi(\vec{r},t)$$

$$+ g \Phi^*(\psi\psi - m) + 2g \Phi(\psi^\dagger \psi - n_T) + g \left( \psi^\dagger \psi \psi - \langle \psi^\dagger \psi \psi \rangle \right)$$

$$+ g_d \left( \int d^3 \vec{r}' n(\vec{r}') U_d(\vec{r},\vec{r}') \right) \psi(\vec{r},t) + g_d \left( \int d^3 \vec{r}' n_c(\vec{r}',\vec{r}) U_d(\vec{r},\vec{r}') \right) \psi(\vec{r}',t)$$

$$+ g_d \left( \int d^3 \vec{r}' m_c(\vec{r}',\vec{r}) U_d(\vec{r},\vec{r}') \psi(\vec{r}',t) \right) + g_{iB} |\varphi|^2 \psi(\vec{r},t)$$

(16)

Near transition temperature, the condensate component is localized around the trap center ($r=0$) and its size is much smaller than the thermal component. So in calculating the energy shift of the thermal component, we can neglect the condensate density $n_c$. From Eq (16)

$$E(\vec{p},\vec{r}) = E^{ide}(\vec{p},\vec{r}) + g_d n_{eff}(\vec{r}) + 2gn(\vec{r}) + g_{iB} n_i(\vec{r}) - \mu \tag{17}$$

where $E^{ide}(\vec{p},\vec{r})$ is the energy spectrum of the ideal Bose gas, $n_{eff}(\vec{r}) = \int d^3\vec{r}' n(\vec{r}') U_d(\vec{r},\vec{r}')$ is an effective density with dipole-dipole interaction, and $n_i(\vec{r}) = |\varphi(\vec{r})|^2$ is the density of impurities.

The transition temperature is determined by the energy spectrum, or more precisely, determined by the shifts of the thermal and condensate states. Thus the main task for the study of the transition temperature is to find the shifts due to the contact, dipole-dipole interactions and impurity.

To determine the energy spectrum of the thermal component consider the phase space Bose-Einstein distribution

$$f(p,r) = \frac{1}{\exp\left[E(p,r)/k_B T\right] - 1} \tag{18}$$

Expand Eq. (17) to the first order of $g$, $g_d$ and $g_{iB}$

$$\begin{aligned} f(p,r) &= f_0(p,r) + g\left(\frac{\partial f(p,r)}{\partial g}\right)_{g=0} + g_d\left(\frac{\partial f(p,r)}{\partial g_d}\right)_{g_d=0} + g_{iB}\left(\frac{\partial f(p,r)}{\partial g_{iB}}\right)_{g_{iB}=0} \\ &= f_0(p,r) - 2gn\frac{\partial f_0(p,r)}{\partial \mu} - g_d n_{eff}(\vec{r})\frac{\partial f_0(p,r)}{\partial \mu} - g_{iB} n_i(r)\frac{\partial f_0(p,r)}{\partial \mu} \end{aligned} \tag{19}$$

where $f_0(p,r)$ is the distribution function of the noninteracting Bose gas.

Integrate over the momentum variables, we get the modification of $n_T(r)$ due to the interactions and impurity

$$n_T^{int}(r) = n_T - 2g n_T \frac{\partial n_T}{\partial \mu} - g_d n_{eff}\frac{\partial n_T}{\partial \mu} - g_{iB} n_i \frac{\partial n_T}{\partial \mu} \tag{20}$$

where $n_T(r)$ is the thermal density distribution of noninteracting Bose gas and in the local density approximation [43] can be written as

$$n_T(r) = \lambda_T^{-3} g_{3/2}\left[\exp\left(U_{ext}(r) - \mu / k_B T\right)\right] \tag{21}$$

here $\lambda_T = \hbar\left(2\pi / mk_B T\right)^{1/2}$ is the thermal wavelength, and $g_\nu(x) = \sum_{n=1}^{\infty} \frac{x^n}{n^\nu}$ is the Bose-Einstein function.

The number of thermal particles is obtained by integrating over the coordinates of Eq. (20)

$$N_T^{int}(r) = \int n_T d^3 r - 2g \int d^3 r n_T \frac{\partial n_T}{\partial \mu} - g_d \int d^3 r n_{eff} \frac{\partial n_T}{\partial \mu} - g_{iB} \int d^3 r n_i \frac{\partial n_T}{\partial \mu} \tag{22}$$

On the other hand, by the Taylor's expansion to the first order of coupling constants

$$N_T^{int}(r) = N_T + g\frac{\partial N_T}{\partial g} + g_d\frac{\partial N_T}{\partial g_d} + g_{iB}\frac{\partial N_T}{\partial g_{iB}} \qquad (23)$$

With our assumption

$$E = E_T^{ide} + g_d\bar{n}_{TT} + 2g\bar{n}_T + g_{iB}\bar{n}_{Ti} - \mu \qquad (24)$$

We get

$$\frac{\partial N_T}{\partial g} = -2\bar{n}_T\frac{\partial N_T}{\partial \mu} \qquad (25)$$

$$\frac{\partial N_T}{\partial g_d} = -\bar{n}_{TT}\frac{\partial N_T}{\partial \mu} \qquad (26)$$

$$\frac{\partial N_T}{\partial g_{iB}} = -\bar{n}_{Ti}\frac{\partial N_T}{\partial \mu} \qquad (27)$$

Put Eqs. (25), (26) and (27) into Eq. (23) and compare with Eq. (22), the effective densities are derived as

$$\bar{n}_T = \frac{\int d^3r \frac{\partial n_T}{\partial \mu} n_T}{\int d^3r \frac{\partial n_T}{\partial \mu}} \qquad (28)$$

$$\bar{n}_{TT} = \frac{\int d^3r \frac{\partial n_T}{\partial \mu} n_{eff}(r)}{\int d^3r \frac{\partial n_T}{\partial \mu}} = \frac{\int d^3r d^3\vec{r}'\left\{\left[\frac{\partial n_T}{\partial \mu}\right]n_T(\vec{r}')U_d(\vec{r},\vec{r}')\right\}}{\int d^3r \frac{\partial n_T}{\partial \mu}} \qquad (29)$$

$$\bar{n}_{Ti} = \frac{\int d^3r \frac{\partial n_T}{\partial \mu} n_i(r)}{\int d^3r \frac{\partial n_T}{\partial \mu}} \qquad (30)$$

where for anisotropic potential $V_{ext}(\vec{r}) = \frac{1}{2}m\omega_z^2 r^2 \Phi(\kappa,\varphi)$

$$\int d^3r \frac{\partial n_T}{\partial \mu} = \frac{2\zeta(2)\kappa^2}{\pi\lambda_T^3 k_B T}\left(\frac{2\pi k_B T}{m\omega^2}\right)^{3/2} \mathrm{E}\left(\sqrt{1-\frac{1}{\kappa^2}}\right) \qquad (31)$$

here E is the second kind Elliptic integral. It is noted that for isotropic case $\left(\kappa = 1 \to \mathrm{E}(0) = \frac{\pi}{2}\right)$ Eq. (31) reduces to Eq. (B5) of ref. [36].

The energy shift of thermal part is found to be

$$\Delta E_T = g_d\bar{n}_{TT} + 2g\bar{n}_T + g_{iB}\bar{n}_{Ti} \qquad (32)$$

At trap center $(r=0), n_T(0) = \xi(3/2)/\lambda_T^3$, Eqs. (28) and (29) give [35, 36]

$$\bar{n}_T = Sn_T(0) \text{ and } \bar{n}_{TT} = -\frac{4\pi}{3}f(\kappa)Sn_T(0)$$

where $S \approx 0.281$ and the anisotropy factor $(f(\kappa))$ is given by [36]

$$f(\kappa) = \begin{cases} \dfrac{2\kappa^2+1}{\kappa^2-1} + \dfrac{3\kappa^2 \operatorname{arcth}\sqrt{1-\kappa^2}}{(1-\kappa^2)^{3/2}} & \kappa \neq 1 \\ 0 & \kappa = 1 \end{cases} \quad (33)$$

The impurity correction to the transition temperature depends on the distribution function of the impurities. For instance, in the case of Gaussian distribution

$$n_i(r) = \frac{1}{(2\pi)^{3/2} \xi^3} e^{-\frac{r^2}{2\xi^2}} \tag{34}$$

where $\xi$ is the correlation length of impurities, Eq. (30) becomes

$$\bar{n}_{Ti} = \left(\frac{m\omega^2}{2k_B T}\right)^{3/2} \frac{2}{(2\pi)^{3/2} \pi \zeta(2) \xi^3 \kappa^2 \mathrm{E}\left(\sqrt{1-\frac{1}{\kappa^2}}\right)}$$

$$\times \sum_{j=1}^{\infty} \frac{1}{j^{1/2}} \left[ \frac{1}{\left(\frac{jm\omega^2}{2k_B T} + \frac{1}{2\xi^2}\right)^{1/2} \left(\frac{jm\omega^2}{2k_B T\kappa^2} + \frac{1}{2\xi^2}\right)} \mathrm{E}\left(\sqrt{\frac{jm\omega^2 \xi^2}{jm\omega^2 \xi^2 + k_B T}\left(1-\frac{1}{\kappa^2}\right)}\right) \right]$$

(35)

The effect of interaction on the condensate can be evaluated by using the eigenfunction of the harmonic oscillator. Following the method of ref [35] the expression of the condensate energy is

$$E = E_c^{ide} + g_d \bar{n}_{cT} + g_d \bar{n}_{cc} + 2gn_T(0) + gn_c + g_{iB}\bar{n}_{ci} - \mu \tag{36}$$

The energy shift of condensate part is

$$\Delta E_c = g_d \bar{n}_{cT} + g_d \bar{n}_{cc} + 2gn_T(0) + gn_c + g_{iB}\bar{n}_{ci} \tag{37}$$

where $E_c^{ide} = \dfrac{3}{2}\hbar\bar{\omega}$ ($\bar{\omega} = \dfrac{\omega_1 + \omega_2 + \omega_3}{3}$) is the energy level of ideal Bose gas in the ground state, $\bar{n}_c = S_c n_c(0) = 0.354 n_c(0)$ and

$$\bar{n}_{cT} = \iint d^3r d^3r' \left[\frac{n_T(\vec{r}) n_c(\vec{r}') U_d(\vec{r}-\vec{r}')}{N_c}\right] \equiv S_{cT} n_T(0) \tag{38}$$

$$\bar{n}_{cc} = \iint d^3r d^3r' \left[\frac{n_c(\vec{r}) n_c(\vec{r}') U_d(\vec{r}-\vec{r}')}{N_c}\right] \equiv S_{cc} n_T(0) \tag{39}$$

$$\bar{n}_{ci} = \int d^3r \left[\frac{n_c(\vec{r}) n_i(\vec{r})}{N_c}\right] \tag{40}$$

here $S_{cT} = S_{cc} = \dfrac{-4\pi}{3} f(\kappa)$ [39].

By using Eq. (34) in to Eq. (40) we have

$$\bar{n}_{ci} = \frac{2}{\pi(2\pi)^{3/2} \kappa \xi^3 a_{ho}^3} \frac{1}{\left(\frac{1}{\kappa^2 a_{ho}^2} + \frac{1}{2\xi^2}\right)\left(\frac{1}{a_{ho}^2} + \frac{1}{2\xi^2}\right)^{1/2}}$$
$$\times E\left[\sqrt{\frac{2\xi^2}{2\xi^2 + a_{ho}^2}\left(1 - \frac{1}{\kappa^2}\right)}\right] \tag{41}$$

The shift of transition temperature due to the energy shift of the thermal and condensate effects can be written as

$$\frac{\delta T_c}{T_c^0} = \frac{\delta T_c^T}{T_c^0} + \frac{\delta T_c^c}{T_c^0} \tag{42}$$

By considering contact, dipole-dipole interactions and impurity contributions, the thermal and condensate parts can be written respectively as

$$\frac{\delta T_c^T}{T_c^0} = \left(\frac{\delta T_c^T}{T_c^0}\right)_{contact} + \left(\frac{\delta T_c^T}{T_c^0}\right)_{dipol-dipol} + \left(\frac{\delta T_c^T}{T_c^0}\right)_{imp} \tag{43}$$

$$\frac{\delta T_c^c}{T_c^0} = \left(\frac{\delta T_c^c}{T_c^0}\right)_{contact} + \left(\frac{\delta T_c^c}{T_c^0}\right)_{dipol-dipol} + \left(\frac{\delta T_c^c}{T_c^0}\right)_{imp} \tag{44}$$

Following the method of ref [35], the shift of the transition temperature due to the energy shifts of the thermal and condensate effects can be calculated as

$$\frac{\delta T_c^T}{T_c^0} = -\frac{\xi(2)}{3[\xi(3)]^{2/3}} \frac{\Delta E_T}{\omega} N^{-1/3}$$
$$= -\frac{\xi(2)}{3[\xi(3)]^{2/3} \omega}\left[(\Delta E_T)_{contact} + (\Delta E_T)_{dipol-dipol} + (\Delta E_T)_{imp}\right] N^{-1/3} \tag{45}$$

$$\frac{\delta T_c^c}{T_c^0} = -\frac{\xi(2)}{3[\xi(3)]^{2/3}} \frac{\Delta E_c}{\omega} N^{-1/3}$$
$$= -\frac{\xi(2)}{3[\xi(3)]^{2/3} \omega}\left[(\Delta E_c)_{contact} + (\Delta E_c)_{dipol-dipol} + (\Delta E_c)_{imp}\right] N^{-1/3} \tag{46}$$

By using Eqs. (32), (35), (37), (38), (39) and (41) into Eqs. (45) and (46) we obtain

$$\frac{\delta T_c^T}{T_c^0} = 2.78 \frac{a_d}{a_{ho}} f(\kappa) N^{1/6} - 1.33 \frac{a}{a_{ho}} N^{1/6} - 3.84 \left(\frac{m\omega^2}{2k_B T}\right)^{3/2} \frac{a_{ho}^2 a_{iB}}{(2\pi)^{3/2} \zeta(2) \xi^3 \kappa^2 E\left(\sqrt{1-\frac{1}{\kappa^2}}\right)}$$
$$\times \sum_{j=1}^{\infty} \frac{1}{j^{1/2}} \left[\frac{1}{\left(\frac{jm\omega^2}{2k_B T} + \frac{1}{2\xi^2}\right)^{1/2} \left(\frac{jm\omega^2}{2k_B T \kappa^2} + \frac{1}{2\xi^2}\right)} E\left(\sqrt{\frac{jm\omega^2 \xi^2}{jm\omega^2 \xi^2 + k_B T}\left(1-\frac{1}{\kappa^2}\right)}\right)\right] N^{-1/3} \tag{47}$$

$$\frac{\delta T_c^c}{T_c^0} = 5.46 \frac{a_d}{a_{ho}} N^{1/6} f(\kappa) - 0.45 \frac{a}{a_{ho}} N^{1/6}$$

$$-3.84 \frac{a_{ho}^2 a_{iB}}{(2\pi)^{3/2} \kappa \xi^3 a_{ho}^3} \frac{1}{\left(\frac{1}{\kappa^2 a_{ho}^2} + \frac{1}{2\xi^2}\right)\left(\frac{1}{a_{ho}^2} + \frac{1}{2\xi^2}\right)^{1/2}}$$

$$\times E\left[\sqrt{\frac{2\xi^2}{2\xi^2 + a_{ho}^2}\left(1 - \frac{1}{\kappa^2}\right)}\right] N^{-1/3} \quad (48)$$

where the effective dipole-dipole scattering length $a_d$ is defined as $g_d = \frac{4\pi a_d \hbar^2}{m}$ and $a_{ho} = \sqrt{\frac{\hbar}{m\omega}}$ is the harmonic oscillator length.

For pancake-shaped potential trap $(\kappa \to \infty)$, and $\xi \Box a_{ho}$, Eqs. (47) and (48) respectively reduce to

$$\frac{\delta T_c^T}{T_c^0} = 5.56 \frac{a_d}{a_{ho}} N^{1/6} - 1.33 \frac{a}{a_{ho}} N^{1/6} - 3.84 \frac{a_{ho}^2 a_{iB}}{(2\pi)^{3/2} \xi^3} N^{-1/3} \quad (49)$$

$$\frac{\delta T_c^c}{T_c^0} = 10.92 \frac{a_d}{a_{ho}} N^{1/6} - 0.45 \frac{a}{a_{ho}} N^{1/6} - 3.84 \frac{a_{ho}^2 a_{iB} \kappa}{(2\pi)^{3/2} \xi^3} N^{-1/3} \quad (50)$$

In this case the total shift of the transition temperature is

$$\frac{\delta T_c}{T_c^0} = 8.24 f(\kappa) \frac{a_d}{a_{ho}} N^{1/6} - 1.78 \frac{a}{a_{ho}} N^{1/6} - 3.84(1+\kappa) \frac{a_{ho}^2 a_{iB}}{(2\pi)^{3/2} \xi^3} N^{-1/3}$$

For isotropic potential trap $(\kappa = 1)$, from Eqs. (47) and (48) we get

$$\frac{\delta T_c^T}{T_c^0} = -1.33 \frac{a}{a_{ho}} N^{1/6} - 3.84 \frac{a_{ho}^2 a_{iB}}{(2\pi)^{3/2} \xi^3} N^{-1/3} \quad (51)$$

$$\frac{\delta T_c^c}{T_c^0} = -0.45 \frac{a}{a_{ho}} N^{1/6} - 0.96 \frac{a_{ho}^2 a_{iB}}{(2\pi)^{1/2} \xi^3} N^{-1/3} \quad (52)$$

The total shift is

$$\frac{\delta T_c}{T_c^0} = -1.78 \frac{a}{a_{ho}} N^{1/6} - 6.02 \frac{a_{ho}^2 a_{iB}}{(2\pi)^{3/2} \xi^3} N^{-1/3} \quad (53)$$

In this case we suggest measuring the impurity effect by tuning the contact scattering length to negligible small by the Feshbach resonance technique.

**3-Coclusion**

The critical temperature of a dipolar gases when the two-particle interaction contains both a short-range, isotropic contact potential and a long-range, anisotropic dipole-dipole interaction in the presence of impurities are considered.

By treating the analytically solvable cases of a Gaussian correlation function, we have estimated that impurity-induced shift of the transition temperature is linear in scattering length and this shift should be positive for boson-impurity attractive interaction $\left(a_{iB}<0\right)$. Since the effect of trap potential is described by the anisotropy factor $\left(f(\kappa)\right)$, to measure the impurity effect on the transition temperature, isotropic trap is used $\left(f(\kappa)=0\right)$ and the contact interaction is tuned to negligible by the Feshbach resonance technique.

**4-References**